\documentclass[%
 aip,
 pof,
 amsmath,amssymb,
 reprint,
 floatfix,
 author-numerical,
]{revtex4-2}

\usepackage[utf8]{inputenc}
\usepackage[T1]{fontenc}
\usepackage{graphicx}
\usepackage{bm}
\usepackage{siunitx}
\usepackage{xcolor}
\usepackage{hyperref}
\graphicspath{{figures/}}

\hypersetup{colorlinks=true, linkcolor=blue, citecolor=blue, urlcolor=blue}
\newcommand{\Dv}{D_{v50}}

\newcommand{\We}{\mathrm{We}}
\newcommand{\Rey}{\mathrm{Re}}
\newcommand{\Fro}{\mathrm{Fr}}
\newcommand{\Oh}{\mathrm{Oh}}

\begin{document}

\title{Weber-number scaling and the distance dependence of drop size in full-cone pressure-swirl sprays}

\author{Bilal Ben Jemaa}
\affiliation{Bioresource Processing Research Institute of Australia (BioPRIA), Department of Chemical and Biological Engineering, Monash University, Clayton, VIC 3800, Australia}
\author{Vincent He}
\affiliation{Venso Pty Ltd, 106 Church Street, Melbourne, Australia}
\author{Marko Trifunovi\'{c}}
\affiliation{Venso Pty Ltd, 106 Church Street, Melbourne, Australia}
\author{Daniel Duke}
\affiliation{Department of Mechanical Engineering, Monash University, Clayton, Australia}
\author{Rishabh More}
\email{Rishabh.More@monash.edu}
\affiliation{Bioresource Processing Research Institute of Australia (BioPRIA), Department of Chemical and Biological Engineering, Monash University, Clayton, VIC 3800, Australia}

\date{\today}

\begin{abstract}
A pressure-swirl nozzle converts the feed pressure into a swirling conical sheet that breaks into ligaments and then into drops. The size of the drops governs the deposition and drift of agricultural sprays, the capture of airborne dust and the combustion efficiency of fuel injectors. The standard tool for measuring the drop size is laser diffraction, which returns a spatial, concentration-weighted distribution at the beam station. However, the dependence of the measured drop size on the nozzle-to-beam distance has not been quantified alongside a Weber-number scaling, so prefactors reported at different distances cannot be compared on an equal footing. To this end, we measure the drop-size distribution of two full-cone pressure-swirl nozzles by laser diffraction at a range of distances, pressures and orientations, together with high-speed imaging of the sheet breakup. We find that the volume-median diameter $\Dv$ decreases with the measurement distance $L$ as $\Dv \sim L^{-0.18}$ beyond the drop-formation region. A joint fit over both nozzles gives $\Dv/d_0 \sim \We^{-0.31}\,(L/d_0)^{-0.21}$, where $d_0$ is the orifice diameter and $\We$ the Weber number, which separates the pressure and distance effects. At a fixed distance the two nozzles collapse onto the $\We^{-1/3}$ scaling of Kooij \textit{et al.}\ (2018), and the measured distance dependence may account for part of the offset between our near-field prefactor and the far-field value. Thus, a drop size obtained by laser diffraction is meaningful only when the measurement distance is reported with it, and a spray should be characterized at the actual distance between the nozzle and its target.
\end{abstract}

\maketitle

\begin{figure*}[tb]
\centering
\includegraphics{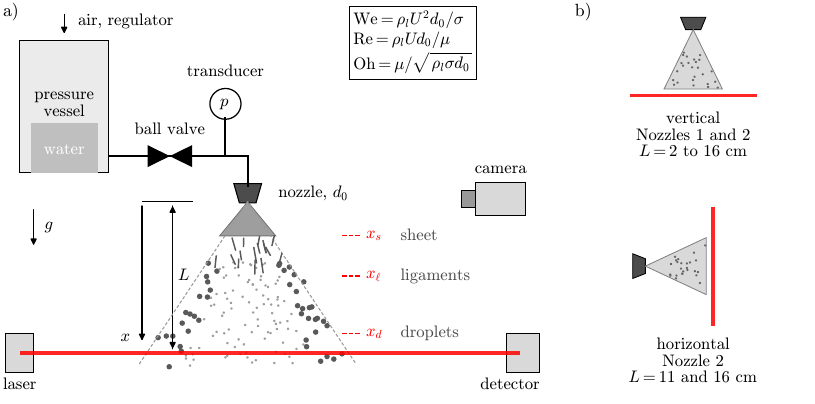}
\caption{(a) Experimental arrangement. Water is fed from a pressurized vessel through a ball valve to the full-cone nozzle of orifice diameter $d_0$. The gauge pressure $p$ is recorded with a transducer, the mean orifice velocity is $U = 4Q/(\pi d_0^2)$ with $Q = K p^{0.43}$, and the dimensionless groups $\We$, $\Rey$ and $\Oh$ are defined in Eq.~(\ref{eq:groups}). The drop-size distribution is measured by laser diffraction with the beam (red line) placed normal to the spray axis at a distance $L$ below the nozzle, and the sheet breakup is imaged with a high-speed camera. The end of the coherent sheet, the last ligament and the start of the drop region are marked $x_s$, $x_\ell$ and $x_d$. (b) The two configurations are the vertical spray (Nozzles~1 and 2) and the horizontal spray (Nozzle~2).}
\label{fig:schematic}
\end{figure*}

\section{Introduction}

Sprays produced by the breakup of a liquid sheet are central to agriculture, dust suppression, drug delivery and gas turbine fuel injection.\cite{LinReitz1998,EggersVillermaux2008,PollockOrganiscak2007,Lefebvre2017} In every case the performance of the process depends on the drop size, whether that is the deposition and drift of an agricultural spray, the capture of airborne dust, or the combustion efficiency of a fuel injector. In each case the fine drops are carried off most easily by the surrounding air, so both the median drop size and the fraction of the liquid carried by fine drops are of practical interest. A pressure-swirl nozzle converts the pressure of the feed liquid into a swirling conical sheet that thins as it expands, becomes unstable, and disintegrates into ligaments and then into drops.\cite{DombrowskiFraser1954,DombrowskiJohns1963,Squire1953} The drop size is therefore set by the competition between the inertia of the sheet and its surface tension, and a large body of work has sought a scaling that predicts the characteristic drop size from the operating conditions and the nozzle geometry.\cite{Villermaux2007,Kooij2018}

The physical picture that underlies the scaling is the ligament-mediated breakup of the sheet. The aerodynamic instability selects a wavelength, the sheet is torn into ligaments, and each ligament collects into drops through capillary instability.\cite{Squire1953,MarmottantVillermaux2004,Villermaux2007} The resulting volume distribution is broad and well described by a gamma form.\cite{MarmottantVillermaux2004,Villermaux2007,Kooij2018} Kooij \textit{et al.}\cite{Kooij2018} showed for flat-fan and hollow-cone nozzles that the volume-median diameter follows $\Dv = C\, b\, \alpha^{-1/6}\,\We_b^{-1/3}$, where $b$ is a sheet-thickness scale, $\alpha = \rho_a/\rho_l$ is the air-to-liquid density ratio, $\We_b = \rho_l U^2 b/\sigma$ is a Weber number built on $b$, the sheet velocity $U$ and the surface tension $\sigma$, and $C \approx 1.95$ is a prefactor of order unity. This scaling has since been confirmed for surfactant and polymer solutions and for Venturi air-induction nozzles.\cite{SijsKooij2021PoF,SijsKooij2023EiF,GaillardSijsBonn2022}

The drop sizes on which these scalings are built are almost always measured by laser diffraction, which is fast and repeatable and has become the standard tool for spray characterization.\cite{DodgeRhodesReitz1987,SijsKooij2021AIP} Other techniques include phase-Doppler particle analysis,\cite{DodgeRhodesReitz1987,Bachalo1984} particle tracking velocimetry,\cite{SijsKooij2021AIP} structured laser illumination planar imaging\cite{Berrocal2008} and X-ray radiography.\cite{KastengrenPowell2014} A laser-diffraction instrument samples the drops present within the beam volume at any instant and therefore returns a spatial, concentration-weighted distribution.\cite{DodgeRhodesReitz1987,SijsKooij2021AIP} When the drop velocity depends on size, as it does once the drops decelerate in still air, the spatially sampled distribution changes along the spray even when no breakup or coalescence takes place.\cite{DodgeRhodesReitz1987} The measurements on which the Kooij \textit{et al.}\ (2018) scaling is built are taken at a single station $40$ to $60$~cm below the nozzle, where the breakup is complete.\cite{Kooij2018,SijsKooij2021PoF,SijsKooij2023EiF} However, the target of a spray often sits much closer to the nozzle, where the drops are younger and faster and the spatial weighting is different. For pressure-swirl fuel injectors, Chin \textit{et al.}\cite{ChinNickolausLefebvre1986} showed analytically that the mean drop size measured by light scattering changes with the distance from the nozzle, because the drops decelerate at rates that depend on their size and the spray disperses radially. For the same reason, Wang and Lefebvre\cite{WangLefebvre1987} placed their diffraction beam $15$~cm from the nozzle as a compromise between this bias and evaporation. For agricultural flat-fan nozzles, Fritz \textit{et al.}\cite{Fritz2014} compared laser diffraction with velocity-weighted imaging at distances from $15$ to $76$~cm. They found that in slow air the diffraction instrument overestimates the volume fraction below $100~\mu$m two- to threefold, and that the bias grows with distance. The downstream evolution of the drop size has also been quantified in swirl sprays by Vankeswaram \textit{et al.}\cite{Vankeswaram2025} using phase-Doppler measurements and at several stations by Peteinaris and Terzis\cite{PeteinarisTerzis2025} for convergent-nozzle sprays. However, the effect of the measurement distance on the apparent drop size has not been quantified alongside a Weber-number scaling, so the prefactors reported at different distances cannot be compared on an equal footing. For a full-cone pressure-swirl nozzle operated at low pressures, neither the Weber-number scaling nor the distance dependence has been established.

To this end, we measure the drop-size distribution of two full-cone pressure-swirl nozzles by laser diffraction at distances from $2$ to $16$~cm, and image the sheet breakup with a high-speed camera, over a range of pressures and orientations. First, we recover the sheet, ligament and drop regions from the imaging and set the near-field limit of the diffraction measurement. Then, we establish the Weber-number scaling at a fixed distance to validate the experimental setup against the Kooij \textit{et al.}\cite{Kooij2018} result. We then quantify the dependence of the measured diameter on distance and orientation, and discuss how much of it comes from the way laser diffraction samples the decelerating fine drops. Finally, we quantify the fine fraction and its dependence on both the Weber number and the working distance, and examine the number-weighted statistics.

\section{Experimental methods}

This section describes the nozzles, the operating conditions and the dimensionless groups, the two measurement techniques and their configurations, and the reproducibility of the data. We consider two commercial full-cone pressure-swirl nozzles (BETE MaxiPass~L, models MPL0.42N and MPL0.21N; BETE, Greenfield, MA) that produce a solid conical spray of water into still air at atmospheric pressure. Nozzle~1 (MPL0.42N) has an orifice diameter $d_0 = 1.30$~mm and Nozzle~2 (MPL0.21N) has $d_0 = 0.94$~mm, with a manufacturer-specified spray cone angle of $51^\circ$. The liquid is deionized water at room temperature, with density $\rho_l = 998$~kg\,m$^{-3}$, surface tension $\sigma = 0.072$~N\,m$^{-1}$ and dynamic viscosity $\mu = 1.0\times10^{-3}$~Pa\,s. The surrounding air has density $\rho_a = 1.2$~kg\,m$^{-3}$, so that the density ratio is $\alpha = \rho_a/\rho_l = 1.2\times10^{-3}$.

\begin{figure}[tb]
\centering
\includegraphics{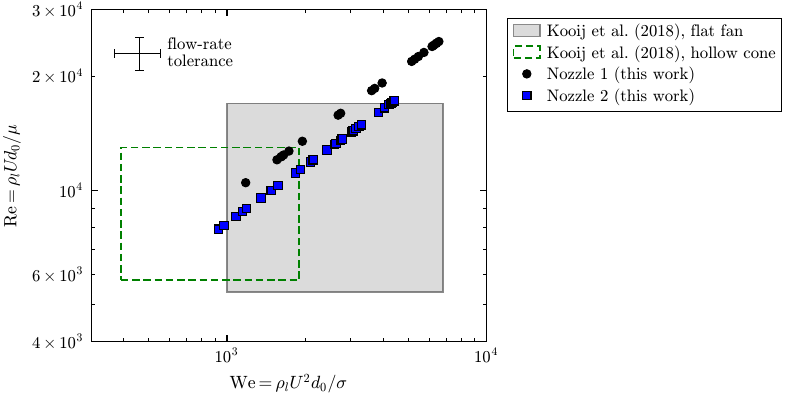}
\caption{Operating range of the present experiments in the Weber--Reynolds plane, with $\We = \rho_l U^2 d_0/\sigma$ and $\Rey = \rho_l U d_0/\mu$ built on the mean orifice velocity $U$ and the orifice diameter $d_0$. Filled markers are the single-nozzle operating points of the present work, i.e., Nozzle~1 (circles) and Nozzle~2 (squares). The grey box and the dashed box are the flat-fan and hollow-cone measurements of Kooij \textit{et al.}\cite{Kooij2018}, recomputed on the same definition from their reported pressures, opening areas and discharge coefficients. The present points sit at $\Rey \sim 10^4$ and $\We \sim 10^3$ to $6.5\times10^3$, deep in the inertia-dominated regime where surface tension, and not viscosity, sets the drop size. The cross at upper left is the representative systematic uncertainty of $\pm 20\%$ in $\We$ and $\pm 10\%$ in $\Rey$, propagated from the $\pm 10\%$ flow-rate tolerance through $U = 4Q/(\pi d_0^2)$.}
\label{fig:regime}
\end{figure}

The liquid is supplied from a pressurized vessel through a ball valve, and the gauge pressure $p$ at the nozzle is recorded with an absolute pressure transducer (Omega PX409) sampled during each run, as shown in Fig.~\ref{fig:schematic}(a). We obtain the gauge pressure as the difference between the plateau mean during spraying and the pre-opening baseline, and the plateau standard deviation is $0.005$ to $0.08$~bar over the range studied. We obtain the volumetric flow rate from the manufacturer relation $Q = K\,p^{0.43}$, with $Q$ in L\,min$^{-1}$ and $p$ in bar, where $K$ is the nozzle flow coefficient, equal to $1.03$ for Nozzle~1 and $0.514$ for Nozzle~2, as given by the manufacturer datasheet (BETE MaxiPass~L series, $\pm 10\%$ flow-rate tolerance). The exponent $0.43$ is below the ideal orifice value of $1/2$, which is consistent with a discharge coefficient that decreases with pressure. The mean orifice velocity is defined as $U = 4Q/(\pi d_0^2)$, and the discharge coefficient $C_d = Q/(A_0\sqrt{2 p/\rho_l})$, with $A_0 = \pi d_0^2/4$ the orifice area, decreases from $0.99$ to $0.82$ over the range studied. The $\pm 10\%$ flow-rate tolerance gives an uncertainty of $\pm 10\%$ in $U$ at each operating point, and hence $\pm 20\%$ in $\We$ and $\pm 10\%$ in $\Rey$, because $\We \propto U^2$ and $\Rey \propto U$. This velocity uncertainty is systematic for a given nozzle, so it shifts the operating points as a group and enters the prefactors of the scalings, and not the fitted exponents. It is shown as a representative error bar in Fig.~\ref{fig:regime}.

We characterize each operating point by the dimensionless groups
\begin{equation}
\We = \frac{\rho_l U^2 d_0}{\sigma}, \qquad
\Rey = \frac{\rho_l U d_0}{\mu}, \qquad
\Oh = \frac{\mu}{\sqrt{\rho_l \sigma d_0}},
\label{eq:groups}
\end{equation}
where $\We$ is the Weber number that compares inertia with surface tension, $\Rey$ is the Reynolds number that compares inertia with viscosity, and $\Oh$ is the Ohnesorge number that compares viscous to capillary-inertial effects. Over the pressures studied we cover $U = 8.1$ to $19.1$~m\,s$^{-1}$, $\We = 0.93\times10^3$ to $6.5\times10^3$ and $\Rey = 7.9\times10^3$ to $2.5\times10^4$. The Ohnesorge number $\Oh = 3.3\times10^{-3}$ to $3.8\times10^{-3}$ is small, so viscous effects on the breakup are weak and the drop size is controlled by the competition between inertia and surface tension. The Froude number $\Fro = U/\sqrt{g d_0} = 72$ to $192$, where $g$ is the gravitational acceleration, so that $\Fro \gg 1$ and gravity is negligible for the atomization itself. These operating points lie at $\Rey \sim 10^4$ and $\We \sim 10^3$, in the inertia-dominated regime of the flat-fan and hollow-cone nozzles for which the Kooij \textit{et al.}\ (2018) scaling was established,\cite{Kooij2018} as shown in Fig.~\ref{fig:regime}.

The operating conditions are summarized in Table~\ref{tab:conditions}.

\begin{table}[tb]
\caption{Operating conditions for the two nozzles. The gauge pressure $p$, mean orifice velocity $U = 4Q/(\pi d_0^2)$ and dimensionless groups are defined in the text. Configuration V is vertical, H horizontal. The number of runs $n$ at each distance $L$ is listed; each run is a separate diffraction record at a distinct pressure. The velocity $U$ carries a systematic $\pm 10\%$ uncertainty from the flow-rate tolerance, i.e., $\pm 20\%$ in $\We$ and $\pm 10\%$ in $\Rey$.}
\label{tab:conditions}
\begin{ruledtabular}
\begin{tabular}{lcccccc}
 & \multicolumn{3}{c}{Nozzle~1 ($d_0 = 1.30$~mm)} & \multicolumn{3}{c}{Nozzle~2 ($d_0 = 0.94$~mm)} \\
\hline
 & Min & Max & $n$ & Min & Max & $n$ \\
\hline
$p$ (bar) & 0.34 & 2.46 & & 0.41 & 2.54 & \\
$U$ (m\,s$^{-1}$) & 8.1 & 19.1 & & 8.4 & 18.4 & \\
$\We$ ($\times 10^3$) & 1.2 & 6.5 & & 0.93 & 4.4 & \\
$\Rey$ ($\times 10^3$) & 10.5 & 25 & & 7.9 & 17 & \\
$\Oh$ ($\times 10^{-3}$) & \multicolumn{2}{c}{3.3} & & \multicolumn{2}{c}{3.8} & \\
$\Fro$ & 72 & 169 & & 88 & 192 & \\
\hline
\multicolumn{7}{l}{\textit{Measurement distances}} \\
$L = 2.1$~cm$^{\mathrm{a)}}$ & & & 4 & & & --- \\
$L = 2.3$~cm$^{\mathrm{a)}}$ & & & --- & & & 3 \\
$L = 3.5$~cm$^{\mathrm{a)}}$ & & & --- & & & 3 \\
$L = 5.5$~cm$^{\mathrm{a)}}$ & & & 4 & & & --- \\
$L = 7.9$~cm & & & --- & & & 13 \\
$L = 8.7$~cm & & & 9 & & & --- \\
$L = 10$~cm & & & 5 & & & --- \\
$L = 11$~cm (H) & & & --- & & & 3 \\
$L = 12.5$~cm & & & --- & & & 4 \\
$L = 16$~cm & & & --- & & & 4 \\
$L = 16$~cm (H) & & & --- & & & 3 \\
\end{tabular}
\end{ruledtabular}
\begin{flushleft}
{\footnotesize $^{\mathrm{a)}}$Inside the drop-formation region ($L < x_d$); excluded from the scaling and distance fits. H denotes horizontal; all others are vertical.}
\end{flushleft}
\end{table}

The drop-size distribution is measured by laser diffraction with a Malvern Spraytec (Malvern Panalytical, 632.8~nm He--Ne laser, 10~mm beam diameter, 300~mm lens, Mie scattering inversion with a refractive index of $1.33$ and zero absorption for water in air, multiple-scattering correction disabled). The drops present in the beam volume scatter the light of the collimated laser and the angular scattering pattern is inverted to a volume-weighted size distribution over $60$ logarithmically spaced bins from $0.1$ to $1000~\mu$m. Most records are sampled at $10$~kHz for $1$~s, and a few at $2.5$~kHz, and the reported distribution is the time average over the record, which does not depend on the sampling rate. The instrument returns the volume-weighted fraction $f_v(d_i)$ in each bin $i$, where $d_i = \sqrt{d_{\mathrm{lo},i}\,d_{\mathrm{hi},i}}$ is the geometric-mean diameter of the bin and $d_{\mathrm{lo},i}$ and $d_{\mathrm{hi},i}$ are its edges. We denote by $\Dv$ the volume-median diameter, by $D_{v10}$ and $D_{v90}$ the $10$th and $90$th volume percentiles, and by $S_v = (D_{v90}-D_{v10})/\Dv$ the relative span of the volume distribution. We obtain the number-weighted fraction $f_n(d_i)$ from the volume fraction by treating the drops as spheres. The number of drops in a bin is its liquid volume divided by the volume $\tfrac{\pi}{6}d_i^3$ of a single drop of that size, so that
\begin{equation}
f_n(d_i) = \frac{f_v(d_i)/d_i^3}{\sum_j f_v(d_j)/d_j^3},
\label{eq:number}
\end{equation}
where the sum runs over the retained bins and normalizes the number distribution to unity. The $1/d_i^3$ weighting amplifies the smallest bins, so the number distribution is dominated by the fine end of the size range, as we discuss in Sec.~\ref{sec:number}. The laser beam is placed normal to the spray axis at a distance $L$ below the nozzle exit, and $L$ is varied from $2.1$ to $16$~cm. The beam passes through the whole spray cone, so that the measurement is a line-of-sight sample across the cone.\cite{DodgeRhodesReitz1987}

\begin{figure*}[tb]
\centering
\includegraphics{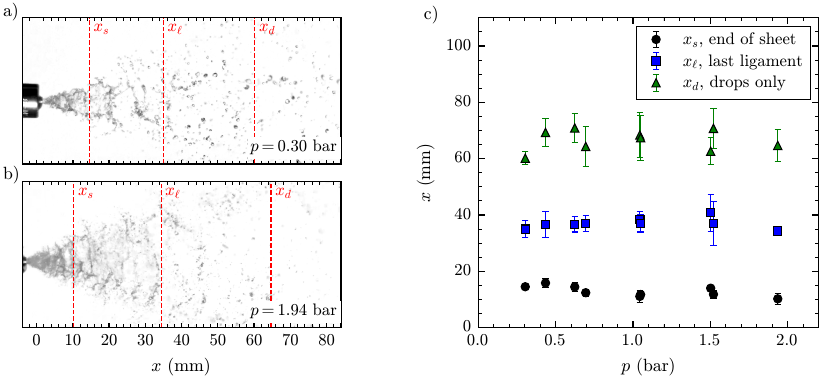}
\caption{High-speed images of the conical sheet of Nozzle~1 at (a) $p=0.30$~bar and (b) $p=1.94$~bar, back-lit, with the axial stations marked by dashed lines: the end of the coherent sheet $x_s$, the last resolved ligament $x_\ell$ and the start of the drop region $x_d$. (c) The three stations against pressure: $x_s$ (circles), $x_\ell$ (squares) and $x_d$ (triangles); the error bars are the standard deviation over the analyzed frames. The sheet ends at $x_s = 10$ to $16$~mm and only drops remain beyond $x_d = 60$ to $71$~mm, so that atomization is complete by $6$ to $7$~cm and the two closest diffraction stations lie inside the ligament region.}
\label{fig:imaging}
\end{figure*}

The sheet breakup of Nozzle~1 is imaged with a high-speed camera (Phantom VEO 410, $5200$~frames\,s$^{-1}$, $1280\times800$~pixels, $\approx 5~\mu$s exposure, $12.4$~pixels\,mm$^{-1}$), back-lit through a diffuser, for gauge pressures from $0.30$ to $1.94$~bar, as shown in Fig.~\ref{fig:imaging}. From each image we extract three axial stations: the end of the coherent sheet $x_s$, the location $x_\ell$ of the last resolved ligament, and the location $x_d$ beyond which only separated drops remain, each measured from the orifice. The stations are identified manually from five frames at each pressure and the reported values are the mean and standard deviation over those frames.

We use two configurations, as shown in Fig.~\ref{fig:schematic}(b). In the vertical configuration the spray points downwards and the beam is placed at a series of distances $L$ below the nozzle. In the horizontal configuration Nozzle~2 sprays sideways and the beam is placed at $L = 11$ and $16$~cm from the exit, which isolates the effect of gravity on the sampling.

The reproducibility of the measurements is quantified from replicate runs and from the within-run scatter. Replicate pairs of runs agree in $\Dv$ to within $0.1$ to $3\%$, with one pair at $9\%$, and the standard deviation of $\Dv$ over five consecutive segments of a single record, i.e., $200$~ms segments for a $1$~s record, is $\sim 1\%$. The within-record segment scatter measures the short-time precision of the diffraction inversion, while the between-run scatter includes the run-to-run variation in pressure, nozzle seating and spray symmetry. The pressure plateau is stable to $0.005$ to $0.08$~bar. We therefore take a representative uncertainty of $\pm 3\%$ on $\Dv$ at a fixed operating point, which includes the between-run reproducibility. The error bars in the scaling and distance plots are the within-record segment standard deviation, which ranges from $1$ to $5\%$ of $\Dv$. The between-run spread, including the $9\%$ outlier pair, is comparable to or larger than this at some operating points. The beam attenuation is below the level at which multiple scattering biases the laser-diffraction size,\cite{DodgeRhodesReitz1987} except at the closest station $L = 2.1$~cm, where the transmission falls to $20$ to $28\%$ for Nozzle~1. We therefore treat the two closest stations, which also lie within the ligament region identified from the imaging, separately from the scaling fits.

\section{Results}

\subsection{Sheet breakup from high-speed imaging}
\label{sec:imaging}

Figure~\ref{fig:imaging} shows the breakup of the conical sheet of Nozzle~1 at the lowest and highest pressures, together with the three axial stations extracted from the images. The coherent sheet ends at $x_s = 10$ to $16$~mm from the orifice, the last resolved ligaments are found at $x_\ell = 34$ to $41$~mm, and only separated drops remain beyond $x_d = 60$ to $71$~mm, as shown in Fig.~\ref{fig:imaging}(c). The end of the sheet moves slightly closer to the orifice as the pressure increases, with $x_s \sim p^{-0.17}$, which is consistent with the aerodynamic sheet instability growing faster at a higher sheet velocity.\cite{Squire1953,DombrowskiFraser1954} The locations $x_\ell$ and $x_d$ show no clear pressure trend within the scatter. The imaging thus places the completion of atomization at $6$ to $7$~cm from the orifice for Nozzle~1. The two closest diffraction stations, at $L = 2.1$ and $2.3$~cm, lie inside the ligament region where the drop population is not yet fully formed, and we exclude every station within the drop-formation length from the scaling and distance analyses. The images show the sheet-to-ligament-to-drop sequence expected for a hydraulic nozzle,\cite{DombrowskiFraser1954,MarmottantVillermaux2004,Villermaux2007} and they set the near-field limit of the diffraction measurement used in the following sections.

\begin{figure}[tb]
\centering
\includegraphics{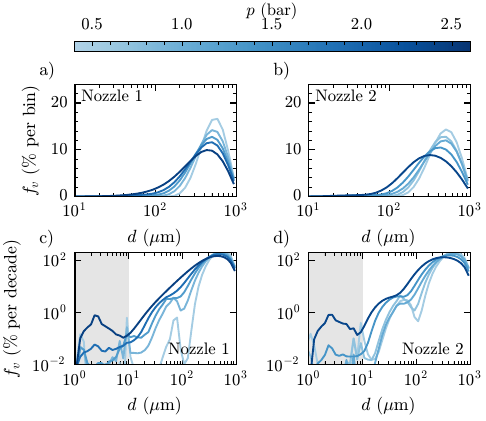}
\caption{Volume-weighted drop-size distributions at a fixed measurement distance for (a,c) Nozzle~1 and (b,d) Nozzle~2, with the gauge pressure $p$ shown by the color shade (light to dark, see colorbar), plotted as (a,b) volume frequency per bin on a linear ordinate and (c,d) volume frequency per decade on logarithmic axes. The distributions are broad and single-peaked, with the peak near $300$ to $600~\mu$m, shift to smaller sizes and broaden as the pressure increases, and show a secondary population at $30$ to $100~\mu$m. The grey sub-$10~\mu$m region is set by the inversion at the instrument limit and is not used in the analysis.}
\label{fig:dists}
\end{figure}

\begin{figure}[tb]
\centering
\includegraphics{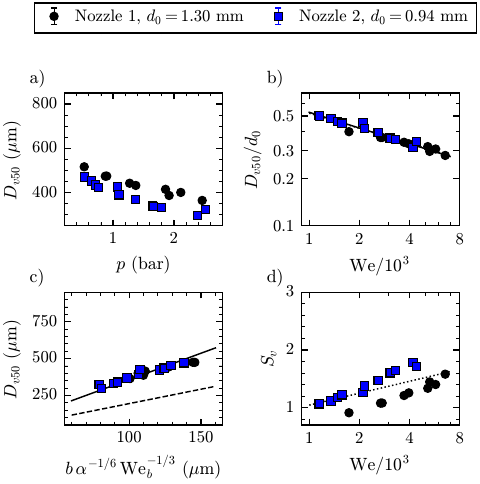}
\caption{Weber-number scaling of the volume-median diameter at a fixed distance for Nozzle~1 (circles) and Nozzle~2 (squares). (a) $\Dv$ against the pressure $p$. (b) $\Dv/d_0$ against $\We$; the solid line is $\Dv/d_0 = 5.3\,\We^{-1/3}$ and the dotted line the free fit $\propto\We^{-0.32}$. (c) $\Dv$ against the Kooij coordinate $b\,\alpha^{-1/6}\We_b^{-1/3}$, with $b=\sqrt{2 C_d A_0/(3\pi)}$ evaluated per operating point; the solid line is the present near-field prefactor $C=3.58$ and the dashed line the value $C=1.95$ measured at $40$ to $60$~cm.\cite{Kooij2018,SijsKooij2021PoF,SijsKooij2023EiF} (d) The relative span $S_v$ against $\We$; the dotted line is $S_v\propto\We^{0.23}$. The error bars are the within-record segment standard deviation and are smaller than the symbols where they are not visible. The two orifice diameters collapse onto a single $\We^{-1/3}$ law, and the near-field prefactor exceeds the far-field value partly because the present data are taken at $8$~cm, as discussed in Sec.~\ref{sec:discussion}.}
\label{fig:scaling}
\end{figure}

\subsection{Drop-size distributions and Weber-number scaling at fixed distance}

Figure~\ref{fig:dists} shows the volume-weighted drop-size distributions of the two nozzles at a fixed measurement distance for a range of pressures. The distributions are broad and single-peaked on a linear ordinate, with the main mode near $300$ to $600~\mu$m and the volume-median near $300$ to $500~\mu$m. The peak shifts to smaller sizes and the distribution broadens as the pressure increases, as shown in Figs.~\ref{fig:dists}(a) and \ref{fig:dists}(b). On logarithmic axes a secondary population appears at $30$ to $100~\mu$m, which carries a small fraction of the volume and grows with pressure, as shown in Figs.~\ref{fig:dists}(c) and \ref{fig:dists}(d). The sub-$10~\mu$m bins carry less than $0.5\%$ of the volume and are shaded, because they are set by the inversion at the limit of the instrument range and are not used in the analysis. The broad, exponentially decaying large-size tail is the signature of ligament-mediated breakup expected for a sheet-based spray.\cite{MarmottantVillermaux2004,Villermaux2007,Kooij2018}

The volume-median diameter decreases with pressure for both nozzles, as shown in Fig.~\ref{fig:scaling}(a), and the larger nozzle produces the larger drops at a given pressure. A fit of both nozzles gives $\Dv \propto p^{-0.26 \pm 0.02}$, which lies within the range of $-0.25$ to $-0.31$ reported for the Sauter mean diameter of low-viscosity liquids from pressure-swirl fuel injectors.\cite{WangLefebvre1987} When the volume-median diameter $\Dv$ is made dimensionless with the orifice diameter $d_0$ and plotted against the Weber number, the data of the two nozzles collapse onto a single power law, as shown in Fig.~\ref{fig:scaling}(b). A free fit of the near-field data (the $22$ vertical runs at $L = 7.9$ and $8.7$~cm) gives an exponent of $-0.32 \pm 0.02$ (standard error from the regression), consistent with the $-1/3$ expected from the inertia-surface-tension competition.\cite{Kooij2018,Villermaux2007} The two distances differ by $10\%$, which under the $L^{-0.18}$ distance dependence of Sec.~\ref{sec:distance} introduces a bias of $1.7\%$ in $\Dv/d_0$, comparable to the difference between the separate prefactors $5.21$ and $5.33$ for the two nozzles. Fixing the exponent at $-1/3$ gives
\begin{equation}
\frac{\Dv}{d_0} = 5.3\,\We^{-1/3},
\label{eq:scaling}
\end{equation}
with a root-mean-square residual of $3.9\%$ and prefactors of $5.21$ and $5.33$ for the two nozzles separately, equal within the scatter. Hence, the two full-cone pressure-swirl nozzles follow the same $\We^{-1/3}$ dependence as the flat-fan and hollow-cone nozzles of Kooij \textit{et al.}\cite{Kooij2018} over the pressures studied, and scaling the drop size with $d_0$ collapses the two nozzles onto one curve.

We compare the prefactor with the value of Kooij \textit{et al.}\cite{Kooij2018} by plotting $\Dv$ against the Kooij \textit{et al.}\ coordinate $b\,\alpha^{-1/6}\,\We_b^{-1/3}$, as shown in Fig.~\ref{fig:scaling}(c). Here the sheet-thickness scale $b = \sqrt{2 C_d A_0/(3\pi)}$ is obtained from the discharge coefficient and the orifice area as in their analysis,\cite{Kooij2018,SijsKooij2023EiF} and the velocity in $\We_b$ is the effective sheet velocity $U_h = Q/(C_d A_0) = U/C_d$ through the contracted orifice area, following Kooij \textit{et al.}, and not the mean orifice velocity $U$. We find $b = 492$ to $527~\mu$m for Nozzle~1 and $347$ to $370~\mu$m for Nozzle~2, and $\alpha^{-1/6} = 3.07$. The near-field data give a prefactor $C = 3.58 \pm 0.24$ (standard deviation over the $22$ runs), which is $1.8$ times the value $C \approx 1.95$ reported for measurements at $40$ to $60$~cm.\cite{Kooij2018,SijsKooij2021PoF,SijsKooij2023EiF} This offset is systematic and is not removed by the choice of $b$. We show in Sec.~\ref{sec:distance} that it is consistent with the smaller distance at which our near-field data are taken.

The relative span of the volume distribution grows with the Weber number as $S_v \sim \We^{0.23}$ over the near-field range, from $0.91$ to $1.78$, as shown in Fig.~\ref{fig:scaling}(d). Thus, the distribution broadens as atomization becomes more vigorous, which is consistent with the ligaments becoming more corrugated at a higher sheet velocity.\cite{Villermaux2007}

\begin{figure}[tb]
\centering
\includegraphics{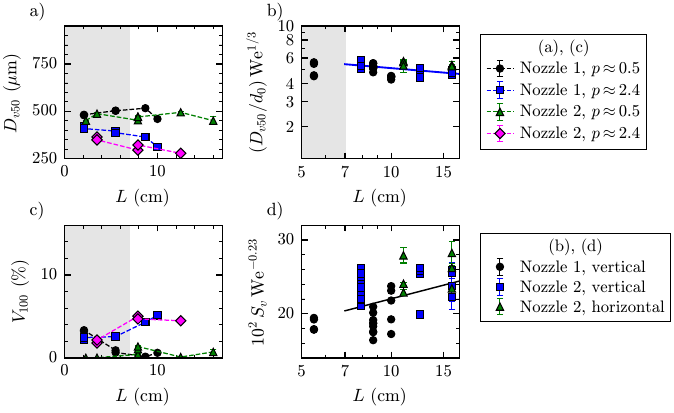}
\caption{Dependence of the measured diameter on distance and orientation. (a) $\Dv$ against the distance $L$ for the two nozzles and gauge pressures $p$ (bar) listed; the shaded region is the drop-formation zone ($L < x_d \approx 7$~cm) excluded from the fits, and the error bars in all panels are the within-record segment standard deviation, and are smaller than the symbols where they are not visible. (b) $(\Dv/d_0)\,\We^{1/3}$ against $L$ for the vertical (circles, squares) and horizontal (triangles) sprays; the line is the Nozzle~2 fit $\propto L^{-0.18}$ over $7.9$ to $16$~cm ($95\%$ confidence interval $[-0.24,-0.11]$, $R^2 = 0.63$). (c) The fine fraction $V_{100}$, i.e., the percentage of the liquid volume below $100~\mu$m, against $L$. (d) $10^2\,S_v\,\We^{-0.23}$ against $L$; the line is $\propto L^{0.20}$. The measured diameter falls with distance and the fine fraction rises, as discussed in Sec.~\ref{sec:distance}.}
\label{fig:distance}
\end{figure}

\subsection{Dependence on measurement distance and orientation}
\label{sec:distance}

Figure~\ref{fig:distance} shows how the measured volume-median diameter depends on the distance $L$ at which the beam is placed. At a fixed pressure the diameter decreases with distance, as shown in Fig.~\ref{fig:distance}(a). We restrict the distance analysis to stations beyond the drop-formation region ($L > x_d \approx 7$~cm) set by the imaging of Sec.~\ref{sec:imaging}, so that the drop population is fully formed. Nozzle~2, which is measured at $L = 7.9$, $12.5$ and $16$~cm, provides the widest range of distances beyond this length. When the near-field scaling of Eq.~(\ref{eq:scaling}) is divided out, the distance dependence is described by
\begin{equation}
\frac{\Dv}{d_0}\,\We^{1/3} \propto L^{-0.18},
\label{eq:distance}
\end{equation}
with a $95\%$ confidence interval of $[-0.24, -0.11]$ on the exponent and a coefficient of determination $R^2 = 0.63$ ($n = 21$ runs at three axial stations), as shown in Fig.~\ref{fig:distance}(b). The beam transmission at these stations is $66$ to $95\%$, above the level at which multiple scattering biases the size,\cite{DodgeRhodesReitz1987} so the trend is not a multiple-scattering artifact. Nozzle~1 is measured beyond the drop-formation length only at $L = 8.7$ and $10$~cm, too short a range to constrain the exponent on its own, and its points are consistent with the Nozzle~2 trend. Hence, the drop size reported for a given nozzle at a given pressure falls by $10$ to $15\%$ between $8$ and $16$~cm. This change is comparable to that produced by a factor of two in pressure, so the prefactor of the Weber-number scaling depends on the distance at which it is measured.

We first consider the residence-time weighting of a spatial line-of-sight measurement as a mechanism for this distance dependence.\cite{DodgeRhodesReitz1987,SijsKooij2021AIP} The laser-diffraction instrument samples the drops present in the beam volume at any instant, so a drop of size $d$ contributes in proportion to the time it spends in the beam, that is, in proportion to $1/u(d)$, where $u(d)$ is its local velocity. Small drops decelerate faster than large drops in still air, so they slow down more with distance and are increasingly over-represented in the spatial sample as the beam is moved away from the nozzle, which lowers the measured median.

We estimate the deceleration of a drop from a drag balance
\begin{equation}
m\,\frac{\mathrm{d}u}{\mathrm{d}t} = -\tfrac12 \rho_a C_D(\Rey_d)\,\frac{\pi d^2}{4}\,u^2,
\label{eq:drag}
\end{equation}
where $m = \rho_l \pi d^3/6$ is the drop mass, $u$ is the drop speed relative to the surrounding air, $d$ is the drop diameter, $C_D$ is the drop drag coefficient and $\Rey_d = \rho_a u d/\mu_a$ is the drop Reynolds number with $\mu_a$ the air viscosity. We take the air to be still, so the relative speed is the drop speed itself, and we integrate from a representative exit speed $u_0 = 15$~m\,s$^{-1}$. The drop Reynolds number is $\Rey_d \approx 30$ to $500$ at this exit speed for drops of $30$ to $500~\mu$m, and it falls as the drop slows. We therefore use the Schiller--Naumann finite-Reynolds-number drag $C_D = (24/\Rey_d)(1 + 0.15\,\Rey_d^{0.687})$ for $\Rey_d < 10^3$, and not the Stokes law, which overestimates the deceleration distances by a factor of $2$ to $6$ over this range.

We characterize the deceleration by the $e$-folding distance $L_e$, the distance over which the drop speed falls to $1/e$ of $u_0$. This gives $L_e \approx 1$~cm for a $30~\mu$m drop, $8$~cm for a $100~\mu$m drop and $40$~cm for a $300~\mu$m drop. Hence, the large drops that set the median travel almost ballistically over the measurement range, while the fine drops lose most of their speed within a few centimeters. In the spray the fine drops relax not to rest but to the slower air entrained by the spray, so the still-air estimate overstates their deceleration. As a result, the fine population is over-weighted increasingly with distance, which lowers the measured median and broadens the distribution, as we observe. This estimate gives the direction of the effect but not its magnitude. Reweighting the measured near-field distribution by the still-air $1/u(d)$ over-predicts the observed median shift, which is consistent with the fine drops relaxing to the entrained air rather than to rest, so the still-air estimate is an upper bound on the sampling contribution. A size-resolved drop velocity $u(d)$ at each station would be needed to predict the magnitude and the exponent.

The drop-scale aerodynamic Weber number $\We_a = \rho_a u^2 d/\sigma$ is $\lesssim 3$ for a $500~\mu$m drop at the highest exit velocity, well below the threshold $\We_a \sim 12$ for aerodynamic breakup.\cite{Villermaux2007} Aerodynamic secondary breakup is therefore unlikely over the measured range. Together with the known concentration weighting of laser diffraction and the size-dependent drag, this makes residence-time weighting a plausible contribution to the observed distance dependence. However, the present line-of-sight measurements cannot separate it from a genuine evolution of the spray, as discussed in Sec.~\ref{sec:discussion}, and do not exclude coalescence. Evaporation, by contrast, is negligible, because the flight time from the breakup region to the beam is only $L/u \sim 10$~ms, over which a water drop at room conditions loses a negligible fraction of its diameter.

The orientation of the spray has a small effect on the measured diameter. The horizontal spray of Nozzle~2, measured at $L = 11$ and $16$~cm, gives a diameter $1.12 \pm 0.03$ times the value of the vertical model extrapolated to the same distance, as shown in Fig.~\ref{fig:distance}(b). Because $\Oh \ll 1$ and $\We \gg 1$, the atomization is controlled by the inertia--surface-tension competition and is not expected to depend on orientation. The difference therefore arises in the post-breakup trajectories and the sampling geometry. In the horizontal spray, gravity deflects the fine drops out of the beam path differently from the vertical spray, so the residence-time weighting changes and the measured median shifts. We do not resolve this further with the present line-of-sight data.

\begin{figure}[tb]
\centering
\includegraphics{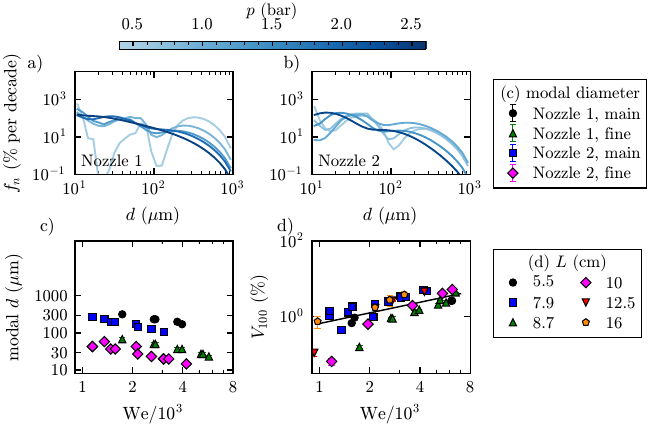}
\caption{Number-weighted statistics and the fine fraction. (a,b) Number-weighted distributions restricted to $d\ge 10~\mu$m for (a) Nozzle~1 and (b) Nozzle~2, with the gauge pressure $p$ shown by the color shade (colorbar). (c) Modal diameters of the number distribution against $\We$: the main mode (circles, squares) at $100$ to $320~\mu$m and the fine mode (triangles, diamonds) at $15$ to $70~\mu$m. (d) The fine fraction $V_{100}$ against $\We$ for the measurement distances $L$ (cm) listed; the line is $\propto\We^{0.94}$ fitted at the reference distance $L \approx 8$~cm only. The number distribution is bimodal above $10~\mu$m, and the fine fraction grows with the Weber number. The error bars in (c) are half the instrument bin width, i.e., $\pm 8\%$, which sets the resolution of a modal diameter, and those in (d) are the within-record segment standard deviation. Error bars smaller than the symbols are not visible.}
\label{fig:number}
\end{figure}

\subsection{Number-weighted statistics and the fine fraction}
\label{sec:number}

The same data can be reweighted by number through Eq.~(\ref{eq:number}). However, the sub-$10~\mu$m bins carry $\le 0.52\%$ of the volume and yet more than $99\%$ of the drop count, so a number-median over the full range is set by the least reliable part of the measurement.\cite{DodgeRhodesReitz1987,SijsKooij2021AIP} As a result, the instrument's own number-median scatters from $3.6$ to $57~\mu$m between replicate runs. We therefore restrict the number distribution to $d \ge 10~\mu$m, below which the volume signal is imposed by the inversion, and report its modes instead of a median, as shown in Figs.~\ref{fig:number}(a) to \ref{fig:number}(c). Above $10~\mu$m the number distribution is bimodal, with a fine mode at $15$ to $70~\mu$m that falls with the Weber number and a main mode at $100$ to $320~\mu$m. This is consistent with the two populations seen in the volume distribution, and with the bimodal near-field distributions reported for swirl and pressure-atomized sprays.\cite{Villermaux2007,Vankeswaram2025,Vallon2021} Thus, the volume-weighted diameter is the reliable output of the measurement, and a number-weighted diameter should not be used without a lower size cut.\cite{SijsKooij2021AIP,Vallon2021}

The fraction of the liquid volume carried by drops below $100~\mu$m, which we denote $V_{100}$, is a practical proxy for the liquid lost to drift, and the same quantity is used to assess the drift potential of agricultural nozzles,\cite{Fritz2014} i.e., the liquid carried by the fine drops that are most easily swept away by the air and miss the target. $V_{100}$ grows with both the Weber number and the distance, as shown in Figs.~\ref{fig:distance}(c) and \ref{fig:number}(d). At the highest pressure $V_{100}$ rises from $2.3\%$ at $L = 2$~cm to $4.5$ to $5\%$ at $L = 8$ to $13$~cm, while at the lowest pressure it stays below $1\%$ beyond $8$~cm. At the reference distance ($L \approx 8$~cm) $V_{100} \sim \We^{0.94}$, as shown in Fig.~\ref{fig:number}(d). Because $V_{100}$ also depends on $L$, this exponent applies at the fixed reference station and should not be extrapolated independently of the distance. $V_{100}$ rises with distance for the same reasons that the median falls, as discussed in Sec.~\ref{sec:distance}, so a fine fraction quoted for a nozzle must be tied to the distance at which it was measured.

\section{Discussion}
\label{sec:discussion}

\begin{figure}[t]
\centering
\includegraphics{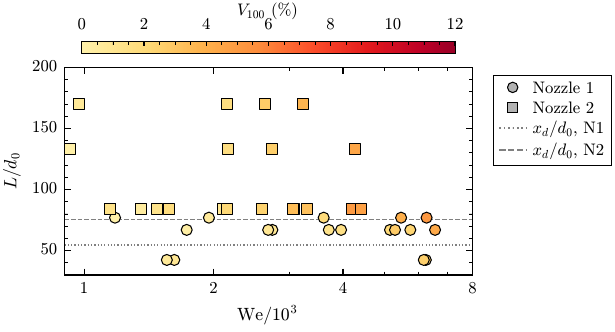}
\caption{Operating map of the measured fine fraction $V_{100}$ over the ($\We$, $L/d_0$) plane for Nozzle~1 (circles) and Nozzle~2 (squares). The horizontal lines mark the drop-formation length $x_d/d_0$ for each nozzle, below which the measurement lies in the ligament region and the laser-diffraction size is unreliable. Higher Weber numbers and larger dimensionless distances both increase the fine fraction. The within-record uncertainty in $V_{100}$ has a median of $5\%$ of its value, i.e., $\sim 0.1$ percentage points.}
\label{fig:phasemap}
\end{figure}

The collapse of two orifice diameters onto $\Dv/d_0 = 5.3\,\We^{-1/3}$ with a residual of a few percent, as shown in Fig.~\ref{fig:scaling}(b), reproduces the $\We^{-1/3}$ dependence of Kooij \textit{et al.}\cite{Kooij2018} for the two full-cone pressure-swirl nozzles and the operating range studied. This collapse is consistent with the drop size being set by the competition between the inertia of the sheet and its surface tension, with $d_0$ as the length scale that collapses the two nozzles. The offset of the prefactor from the value of Kooij \textit{et al.}\cite{Kooij2018,SijsKooij2021PoF,SijsKooij2023EiF} has more than one likely cause, and the present data cannot distinguish between them. One cause is the different measurement distance, i.e., our near-field data at $8$~cm against their far-field data at $40$ to $60$~cm. If the measured $L^{-0.18}$ dependence of Eq.~(\ref{eq:distance}) continued out to $40$ to $60$~cm, it would lower our near-field prefactor by a factor of $1.3$ to $1.4$, i.e., from $3.58$ to $\sim 2.6$, which removes part of the difference from $1.95$ but leaves a clear residual. A second, and possibly larger, cause is that our $\Dv$ is a line-of-sight average across the whole solid cone, which mixes the coarse core and the finer edge, and so is not the same quantity as a local drop size. The full-cone geometry and the definition of the sheet-thickness scale $b$, which enters the prefactor directly, may also contribute. We cannot separate these contributions with the present data.

A joint fit of the vertical single-nozzle data beyond the drop-formation length ($n = 35$ runs at $L = 7.9$ to $16$~cm) to $\log(\Dv/d_0) = A + m\log\We + n\log(L/d_0) + \eta\,I_{\mathrm{N2}}$, where $I_{\mathrm{N2}}$ is an indicator for Nozzle~2, gives $m = -0.31 \pm 0.02$, $n = -0.21 \pm 0.04$ and $\eta = 0.14 \pm 0.03$ ($R^2 = 0.92$). In dimensional form this is $\Dv/d_0 = 9.4\,\We^{-0.31}\,(L/d_0)^{-0.21}$, with Nozzle~2 $15\%$ higher than Nozzle~1 at the same $\We$ and $L/d_0$. The Weber exponent is consistent with the $-1/3$ of the fixed-distance fit and the distance exponent is consistent with the separate $L^{-0.18}$ of Eq.~(\ref{eq:distance}), so the two effects are additive in the logarithm and do not interact over the range studied.

The measured distance dependence has two contributions that our line-of-sight data cannot separate. First, laser diffraction returns a spatial, concentration-weighted distribution, in which each drop is weighted by the time it spends in the beam, so the measured distribution changes along the spray when the drop velocity depends on size.\cite{DodgeRhodesReitz1987,SijsKooij2021AIP} Chin \textit{et al.}\cite{ChinNickolausLefebvre1986} computed this effect for pressure-swirl sprays. They showed that, when the drops leave the nozzle faster than the surrounding air, the measured mass median diameter first falls with distance and later recovers, and that the change persists farther downstream for coarser sprays. For accelerating drops they found maximum deviations of $3$ to $12\%$, larger for broader distributions. Our sprays are coarse and broad and decelerate in still air, so the $10$ to $15\%$ fall that we measure between $8$ and $16$~cm is consistent with the first branch of their prediction and comparable in size. For flat-fan nozzles in air at $0.7$~m\,s$^{-1}$, Fritz \textit{et al.}\cite{Fritz2014} measured the same trend, with the laser-diffraction median of their two coarser nozzles falling by about $18$ to $19\%$ between $15$ and $76$~cm. By comparing with velocity-weighted imaging at the same stations, they showed that the sampling bias alone grows from a few percent at $15$~cm to over $20\%$ at $76$~cm, and that a coflowing air stream close to the drop velocity reduces it to a few percent. Second, the spray itself evolves. Phase-Doppler measurements of a swirl spray show that beyond about twice the film-breakup length the drops are decelerated and sorted by the surrounding air, with the fine drops carried by the air and the large drops remaining ballistic.\cite{Vankeswaram2025} Both effects lower the measured median with distance, and a size-resolved velocity measurement would be needed to distinguish them. Whatever the split, a drop size or a fine fraction obtained by laser diffraction is meaningful only when the measurement distance is reported with it. In practice, the median drop size and the fine fraction should be characterized at the actual distance between the nozzle and its target, and a drop size quoted at a different distance will misstate both.

We summarize the combined effect of the Weber number and the working distance on the fine fraction in an operating map, shown in Fig.~\ref{fig:phasemap}. The fraction of liquid volume carried by drops below $100~\mu$m is a practical proxy for the propensity of a spray to drift, and it depends on both the pressure and the distance between the nozzle and the target. The loss itself also depends on the drop velocity, the airflow around the target, the target geometry and the capture efficiency, which we do not measure here. The operating map shows $V_{100}$ over the ($\We$, $L/d_0$) plane for both nozzles, from which the trend of the fine fraction with pressure and working distance can be seen. The map confirms that the fine fraction grows with the Weber number and with the dimensionless distance, as expected from the scaling of Sec.~\ref{sec:distance}, and that measurements taken in the ligament region, below $x_d/d_0$, should not be used for this purpose.

\section{Conclusions}

In this study, we have measured the drop-size distribution of two full-cone pressure-swirl nozzles by laser diffraction at distances from $2$ to $16$~cm, together with high-speed imaging of the sheet breakup, over a range of pressures and orientations. We find that the volume-median diameter measured by laser diffraction decreases with distance as $\Dv \sim L^{-0.18}$ beyond the drop-formation region, so that the apparent drop size at $16$~cm is $10$ to $15\%$ smaller than at $8$~cm. We interpret this decrease as arising partly from the residence-time weighting of a spatial measurement, in which fine drops decelerate faster than large drops and are over-represented increasingly far from the nozzle, and partly from a genuine spatial evolution of the spray that our line-of-sight data cannot separate from it. A joint fit over both nozzles gives $\Dv/d_0 = 9.4\,\We^{-0.31}\,(L/d_0)^{-0.21}$, which separates the pressure and distance effects. At a fixed distance the two nozzles collapse onto $\Dv/d_0 = 5.3\,\We^{-1/3}$, the same Weber-number dependence found by Kooij \textit{et al.}\ (2018) for flat-fan and hollow-cone nozzles, which validates the experimental approach. The near-field prefactor exceeds the far-field value by a factor of $1.8$. The measured distance dependence may account for part of this offset, with the remainder likely due to the full-cone geometry and the whole-cone line-of-sight average. A drop size obtained by laser diffraction is therefore meaningful only when the measurement distance is reported with it. A spray should therefore be characterized at the actual distance between the nozzle and its target, and the fine-fraction data and operating map of this study can be used to select the pressure and working distance that limit the fine fraction. The present measurements use water. Many agricultural and industrial sprays carry suspended fibers or particles, whose shear-thinning viscosity depends on the volume fraction, the stress and the particle aspect ratio.\cite{KhanMoreArdekani2023} Extending the distance-resolved measurements to such suspensions is a natural next step.

\begin{acknowledgments}
The authors acknowledge SAMPLE Labs at Monash University for technical support. R.V.M. acknowledges support from the Engineering Talent Research Accelerator (ERTA) Fellowship (2025) from the Faculty of Engineering at Monash University. This research was supported by the Industry Innovation Program at Monash University. D.D. acknowledges the Faculty of Engineering Lab Enhancement Program (2025) and ERTA (2026) at Monash University for financially supporting the laboratory equipment used in this research.
\end{acknowledgments}

\section*{Author declarations}
\subsection*{Conflict of interest}
The authors have no conflicts to disclose.

\subsection*{Author contributions}
\textbf{Bilal Ben Jemaa}: Investigation; Methodology; Formal analysis; Data curation; Visualization; Writing -- original draft. \textbf{Vincent He} and \textbf{Marko Trifunovi\'{c}}: Methodology; Writing -- review and editing. \textbf{Daniel Duke}: Methodology; Writing -- review and editing. \textbf{Rishabh More}: Supervision; Conceptualization; Methodology; Investigation; Formal analysis; Writing -- original draft, review and editing.

\section*{Data availability}
The data that support the findings of this study are available from the corresponding author upon reasonable request.

\end{document}